\documentclass[fleqn,12pt]{wlscirep}

\usepackage{hyperref}
\usepackage{amssymb}
\usepackage[utf8]{inputenc}
\usepackage{multirow}
\usepackage{times}
\usepackage{chngcntr}
\usepackage[normalem]{ulem}
\usepackage{float}
\usepackage{lineno}

\usepackage{multibib}
\newcites{online}{Additional references}

\newcommand\figref[1]{Fig.\,\ref{#1}}
\newcommand\edfigref[1]{Extended Data Fig.\,\ref{#1}}

\newcommand\equref[1]{Eq.\,\eqref{#1}}

\newcommand\edtabref[1]{Table\,\ref{#1}}

\newcommand{\kpc}{{\mathrm{kpc}}}

\newcommand{\Gaia}{{\it Gaia }}
\newcommand{\Gaiabf}{{\bf{\textit{Gaia }}}}

\newcounter{firstbib}

\def\be{\begin{equation}}
\def\ee{\end{equation}}
\def\kms2{{\rm\,(km\,s^{-1})^2}}
\def\kms{{\rm\,km\,s^{-1}}}
\def\kmskpc{{\rm\,km\,s^{-1}\,{kpc}^{-1}}}

\def\Myr{{\rm\,Myr}}
\def\Gyr{{\rm\,Gyr}}
\def\deg{{^\circ}}

\def\kpc{{\rm\,kpc}}

\def\1s{{1$\sigma$}}
\def\2s{{2$\sigma$}}
\def\3s{{3$\sigma$}}

\def\Vr{V_{{R}}}
\def\Vp{V_{{\phi}}}
\def\VZ{V_{{Z}}}
\def\b#1{{\bf #1}}

\def\mnras{Mon. Not. R. Astron. Soc.}
\def\aap{Astron. Astrophys.}
\def\apj{Astrophys. J.}
\def\apjl{Astrophys. J.}
\def\aj{Astron. J.}

\def\pasj{PASJ}

\def\rmxaa{Rev. Mexicana Astron. Astrofis.}
\def\nat{Nature}

\title{A dynamically young and perturbed Milky Way disk}

\author[1]{T. Antoja}
\author[2]{A. Helmi}
\author[1]{M. Romero-G\'{o}mez}
\author[3]{D. Katz}
\author[4]{C. Babusiaux}
\author[5]{R. Drimmel}
\author[6]{D. W. Evans}
\author[1]{F. Figueras}
\author[5,7]{E. Poggio}
\author[8]{C. Reyl\'{e}}
\author[8]{A.C. Robin}
\author[9]{G. Seabroke}
\author[10]{C. Soubiran}

\affil[ ]{}

\begin{abstract}

\b{Most of the stars in our Galaxy including our Sun, move in a disk-like component\cite{Oort1927} and give the Milky Way its characteristic appearance on the night sky.  
As in all fields in science, motions can be used to reveal the underlying forces, and in the case of disk stars they provide important diagnostics on the structure and history of the Galaxy\cite{Bible2008}. But because of the challenges involved in measuring stellar motions, samples have so far
 remained limited in their
 number of stars, precision and spatial extent. 
This has changed dramatically with the second Data Release of the \Gaiabf mission\cite{Brown2018} which has just become available. 
Here we report that the phase space distribution of stars in the disk of the Milky Way is full of substructure with a variety of morphologies  
never 
 observed before, \b{namely}  
 snail shells and ridges when spatial and velocity coordinates are combined.  
The nature of these \b{new} substructures implies that the disk is phase mixing from an out of equilibrium state, and that  
\b{it is strongly affected by the Galactic bar and/or spiral structure.} 
 Our analysis of the features
leads us to infer that the disk was perturbed between 300 and 900 Myr ago, matching current estimations of the previous pericentric passage of the Sagittarius dwarf galaxy. 
\b{Although phase-wrapping was predicted to occur in the disk after a passage of a satellite galaxy\cite{Minchev2009,Gomez2012a,delaVega2015}, the substructures discovered here were never examined before.}  
The \Gaiabf data challenge the most basic premise \b{in} 
stellar dynamics of dynamical equilibrium, and show that modelling the Galactic disk as a time-independent axisymmetric component is definitively incorrect. 
These findings mark a new era when, by modelling the richness of phase space substructures, we can determine the structure of the Galaxy and the characteristics of the perturbers that have most influenced our home in the Universe. } 

\end{abstract}
\begin{document}

\flushbottom
\maketitle
%
%
\thispagestyle{empty}


\newpage

\section*{Main}



The evolution of the disk of our Galaxy has been sculpted by several phenomena whose exact roles have yet to be constrained. Models have shown that the bar and the spiral arms can have a major impact on the disk dynamics, for instance inducing 
radial migration\cite{Sellwood2002} and trapping/scattering close to orbital resonances\cite{Contopoulos1986}. External perturbations from satellite galaxies must also play a role, causing for example, heating\cite{Quinn1993},  rings\cite{Purcell2011}, correlations between velocities\cite{DOnghia2015}, and phase-wrapping signatures 
 in the disk\cite{Minchev2009,Gomez2012a,delaVega2015}, such as arched velocity structures in the motions of stars in the Galactic plane.
Some manifestations of these dynamical processes have been already detected in observations. These include mostly kinematic substructure in samples of nearby stars \cite{Eggen1996,Dehnen1998,Katz2018}, 
 density asymmetries and velocities across the Galaxy disk that differ from the axisymmetric and equilibrium expectations\cite{Siebert2011a}, especially in the vertical direction\cite{Dehnen1998,Widrow2012,Schonrich2017,Quillen2018}, 
  and signatures of incomplete phase-mixing in the disk\cite{Minchev2009,Gomez2012b,Monari2018,Katz2018}.

\Gaia is a cornerstone mission of the European Space Agency (ESA) that has been designed primarily to investigate  the origin, evolution and structure of the Milky Way and has just delivered an exquisite product: the largest and most precise census of positions, velocities and other stellar properties for more than a billion stars. By exploring the phase space of more than 6 million stars (positions and velocities) in the disk of the Galaxy in the first kiloparsecs around the Sun from the \Gaia Data Release 2 (DR2, see Methods), we find that certain phase space projections show plenty of substructures that are new and that had not been predicted by existing models. These have remained blurred until now due to the limitations on the number of stars and the precision of the previously available datasets.
\figref{fig:ZVZ}a shows the projection of phase space in the vertical position and velocity $Z$-$\VZ$. The stars follow an impressive curled spiral-shaped distribution never seen before whose density increases towards the leading edge of the spiral. 
\figref{fig:ZVZ}b and \figref{fig:ZVZ}c show that the ``snail shell''  
is still present when the stars are colour-coded according to their $\Vr$ and $\Vp$ values, implying a strong correlation between the vertical and in-plane motions of the stars. The pattern is particularly pronounced in the $\Vp$ colour-coded case (\figref{fig:ZVZ}c) even  up to $\VZ\sim40\kms$. Furthermore, we see a gradient 
with 
different azimuthal velocities across the spiral shape, following the density variations.
Details about the relation between the ``snail shell'' and other velocity features observed in the Solar neighbourhood are described in \edfigref{fig:VrVphiVzSN}.


The spiral shape of \figref{fig:ZVZ}a is clearly reminiscent of the effects of phase mixing in two-dimensions discussed in several areas of Astrophysics\cite{Tremaine1999,Afshordi2008,Candlish2013} 
and also in quantum physics\cite{Manfredi1996} but never put in the context of dynamical models of the disk. 
This process can be better understood with a simple toy model. Consider a Galaxy model 
 whose vertical potential can be approximated by an anharmonic oscillator of the shape
 \begin{equation}\label{anharmonic}
 \Phi(Z)\propto -\alpha_0 +\frac{1}{2}\alpha_1Z^2-\frac{1}{4}\alpha_2Z^4,
 \end{equation}
\noindent where the coefficients $\alpha_i$ depend on Galactocentric radius $R$, since the vertical pull depends on the distance to the Galactic center. In this approximation, the frequencies of oscillation depend on the amplitude of the oscillation $A$, to first order, as\cite{Candlish2013}
\begin{equation}\label{frequencies}
\nu(A,R)=\alpha_1(R)^{1/2}\left(1-\frac{3\alpha_2(R)A^2}{8\alpha_1(R)}\right)
\end{equation}
\noindent { where $\nu_0 \equiv\alpha_1^{1/2}$ is the vertical frequency in the epicyclic approximation. By approximating that stars follow a simple harmonic oscillation with these frequencies, their movement with time is described by
\begin{equation}\label{harmonic}
Z=A\cos\left(\nu(A)t+\phi_0\right), \hspace{1cm}V_Z=-A\nu(A)\sin\left(\nu(A)t+\phi_0\right),
\end{equation}
\noindent  which traces a circle in the clockwise direction in the $Z$-$V_Z$ projection. However, stars revolve at different angular speeds depending on their frequency. Thus, an ensemble of stars will stretch out in phase space, with the range of frequencies causing a spiral shape in this projection. The detailed time evolution of stars in this toy model is described in Methods and shown in \edfigref{fig:snailmodel}.}  
As time goes by, the spiral gets more tightly wound, and eventually, this process of phase mixing leads to a spiral that is so wound that the coarse-grained distribution appears to be smooth. The clarity of the spiral shape in the $Z$-$\VZ$ plane revealed by the \Gaia DR2 data, implies that this time has not yet arrived and thus provides unique evidence that phase mixing is currently taking place in the disk of the Galaxy. 


This interpretation also implies that the shape of the spiral can be used to obtain information on: i) the shape of the potential, which determines the vertical frequencies, ii) the starting time of the phase mixing, and iii) the type of perturbation that brought the disk to a non-equilibrium state, which sets the initial conditions for the phase mixing event we are witnessing. For instance, we can estimate the time $t$ of the event from the separation between two consecutive spiral turns because these have phase separation of $2\pi$, that is $(\nu_2 t+\phi_{0_2}) - (\nu_1 t+\phi_{0_1})=2\pi$, where 1 and 2 indicate two consecutive turns of the spiral. Therefore, we have that 
\begin{equation}\label{time}
t=\frac{2\pi}{\nu_2-\nu_1},
\end{equation}
\noindent where we have assumed that the initial phase is the same for 1 and 2. Using several potentials for the Milky Way, as explained in Methods, we estimated that the vertical phase mixing event started about 500 Myr ago, with a likely range of $[300, 900 \Myr]$. A toy model that illustrates this process is shown in \figref{fig:models}a, which depicts a spiral shape similar to the data formed after 500 Myr from an ensemble of stars with a starting distribution that is out of equilibrium (see Methods). 

A possible perturbation that might have sparked the observed on-going vertical phase mixing is the influence of a satellite galaxy. In particular, the last pericentre of the orbit of the Sagittarius dwarf galaxy has been shown to have strong effects on the stellar disk \cite{Purcell2011,Gomez2012a,delaVega2015}. 
In addition, most models locate this pericentric passage between 200 and 1000 Myr ago\cite{Law2010,delaVega2015,Laporte2017}, which is fully consistent with our findings. Nevertheless, other processes that may induce snail shells could be the formation of the central bar and of transient spiral structure, provided that these are able to induce vertical asymmetries, other global changes in the potential, 
or the dissolution of a massive stellar system such as a cluster or accreted satellite. 

\begin{figure}[H]
\centering
\includegraphics[width=0.6\linewidth]{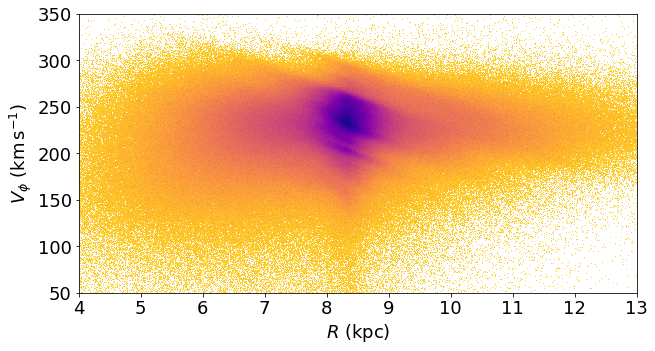}

\caption{{\bf Distribution of azimuthal velocities as a function of Galactocentric radius from \Gaiabf DR2 data.} Two-dimensional histogram for all observed stars in our sample with 6D phase space coordinates in bins of $\Delta\Vp=1.\kms$, and $\Delta R=0.01\kpc$. $\Vp$ is positive towards the Galactic rotation direction. 
}
\label{fig:VphiRphiZ}
\end{figure}

Another phase space projection that shows a remarkably different and stunning appearance with the Gaia data is the azimuthal velocity $\Vp$ versus cylindric radius $R$ (\figref{fig:VphiRphiZ}). Although this phase space projection was explored before with other data\cite{Monari2017}, the spatial coverage, high sampling and unprecedented precision of the \Gaia data unveils a plethora of diagonal thin ridges. 
 The arches in the velocity space projection $\Vr$-$\Vp$ at the Solar neighbourhood recently discovered with \Gaia data\cite{Katz2018} and shown in \edfigref{fig:VrVphiVzSN}a are projections of these diagonal ridges but at a fixed Galactic position. \figref{fig:VphiRphiZ} this reveals that arched structures must be present at many different radii but have remained fully unexplored thus far,  and that their characteristics vary with distance from the Galactic centre, diminishing their velocity towards the outskirts of the Galaxy in a continuous way.

These diagonal ridges 
 could be signatures of phase mixing now in the horizontal direction, as has been already predicted for the arches in velocity space\cite{Fux2001,Minchev2009,Gomez2012a}. 
Alternatively, the bar and the spiral arms could also induce diagonal ridges  created by the resonant orbital structure,  
 which exhibit regions in phase space of 
  stable and unstable orbits\cite{Michtchenko2018}, and hence with over-densities and gaps. %
A simple toy model of phase mixing currently at work (\figref{fig:models}b) but also, a disk simulation with a Galactic potential containing a bar (\figref{fig:models}c)  
show plenty of diagonal ridges. 
 The $\Vp$ velocity separation of consecutive ridges in the data is about $10\kms$.
This separation compared to that of all our toy models (see Methods) seems to indicate that if these ridges were caused by phase mixing from a single perturbation, this should have taken place a longer time ago than the one giving rise to the vertical mixing, 
This conclusion is consistent with the timing derived using the separation between arches in the local velocity plane\cite{Minchev2009} and some stellar moving groups that appear not to be fully phase mixed vertically\cite{Monari2018}, which was found to be 2 Gyr ago. The relation between the various features is not clear, and it is not unlikely that  there are/were several perturbations creating superposed features.

We have here provided the clearest evidence hitherto that our own Galaxy disk has suffered from perturbations bringing it to an out of equilibrium state, which may well be due to the interaction with an external satellite galaxy. Our findings indicate that the emergent field of galactoseismology\cite{Widrow2012} will become a consolidated reality in the Gaia era.
Our interpretation of the new features found, however, is based on simple toy models, with their main limitations being the lack of self-consistency, the choice of initial conditions not necessarily reflecting those stemming from the impact of a satellite galaxy, and the fact that we study separately the effects of resonances and phase mixing and also the different dimensions (horizontal and vertical) at play. A challenging task for the future will be to model the new findings taking into account collective effects, such as in the perturbative regime\cite{Fouvry2015} and also with self-consistent N-body models\cite{Laporte2017}.

\renewcommand{\refname}{References}

\section*{Acknowledgements}


This work has made use of data from the European Space Agency (ESA) mission {\it Gaia} (\url{https://www.cosmos.esa.int/gaia}), processed by the {\it Gaia} Data Processing and Analysis Consortium (DPAC, \url{https://www.cosmos.esa.int/web/gaia/dpac/consortium}). Funding for the DPAC has been provided by national institutions, in particular the institutions participating in the {\it Gaia} Multilateral Agreement. 
This project has received funding from the European Union's Horizon 2020 research and innovation
programme under the Marie Sk{\l}odowska-Curie grant agreement No. 745617. 
This work was supported by the 
MDM-2014-0369 of ICCUB (Unidad de Excelencia 'Mar\'\i a de Maeztu') and the European Community's Seventh Framework Programme (FP7/2007-2013) under grant agreement GENIUS FP7 - 606740. AH acknowledges financial support from a VICI grant from the Netherlands Organisation for Scientific Research, NWO. We acknowledge the 
MINECO (Spanish Ministry of Economy) through grants ESP2016-80079-C2-1-R (MINECO/FEDER, UE) and ESP2014-55996-C2-1-R (MINECO/FEDER, UE).

\section*{Author information}

\subsection*{Affiliations}

\noindent $^1${Institut de Ci\`{e}ncies del Cosmos, Universitat  de  Barcelona  (IEEC-UB), Mart\'{i} i Franqu\`{e}s  1, E-08028 Barcelona, Spain}

\noindent $^2${Kapteyn Astronomical Institute, University of Groningen, Landleven 12, 9747 AD Groningen, The Netherlands}

\noindent $^3${GEPI, Observatoire de Paris, Universit\'{e} PSL, CNRS, 5 Place Jules Janssen, 92190 Meudon, France}

\noindent $^4${Univ. Grenoble Alpes, CNRS, IPAG, 38000 Grenoble, France}

\noindent $^5${INAF - Osservatorio Astrofisico di Torino, via Osservatorio 20, 10025 Pino Torinese (TO), Italy}

\noindent $^6${Institute of Astronomy, University of Cambridge, Madingley Road, Cambridge CB3 0HA, UK.}

\noindent $^7${Universit\`{a} di Torino, Dipartimento di Fisica, via Pietro Giuria 1, 10125 Torino, Italy}

\noindent $^8${Institut UTINAM, CNRS UMR6213, Univ. Bourgogne Franche-Comt\'e, OSU THETA Franche-Comt\'e Bourgogne, Observatoire de Besan\c con, BP 1615, 25010 Besan\c con Cedex, France. }

\noindent $^9${Mullard Space Science Laboratory, University College London, Holmbury St Mary, Dorking, Surrey RH5 6NT, United Kingdom}

\noindent $^{10}${Laboratoire d'astrophysique de Bordeaux, Univ. Bordeaux, CNRS, B18N, all{\'e}e Geoffroy Saint-Hilaire, 33615 Pessac, France}

\subsection*{Contributions statement}
T.A. contributed to the sample preparation, analysed and interpreted the data, performed most of the modelling and wrote the paper together 
with A.H. A.H. provided interpretation to the findings. M.R. performed the simulation with the barred potential and contributed to sample preparation. D.K., C.B, R.D., D.W.E, F.F., E.P., C.R., A.C.R, G.S, C.S. contributed to sample preparation and validation of the \Gaia data.
All authors reviewed the manuscript. 

\subsection*{Corresponding author and request for materials}

Correspondence to \href{mailto:tantoja@fqa.ub.edu}{Teresa Antoja}.

\subsection*{Competing interests}
The authors declare no competing financial interests.




\begin{figure}[H]
\centering
\includegraphics[height=0.19\textheight]{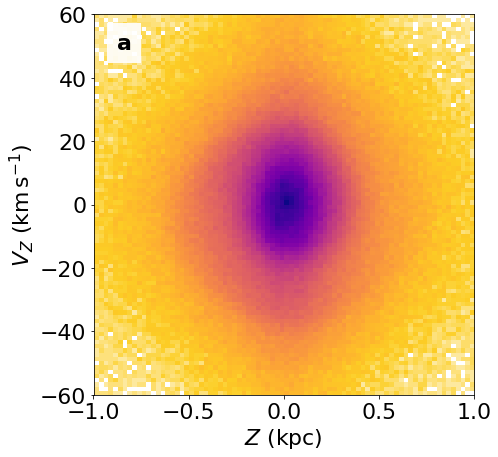}
\includegraphics[height=0.19\textheight]{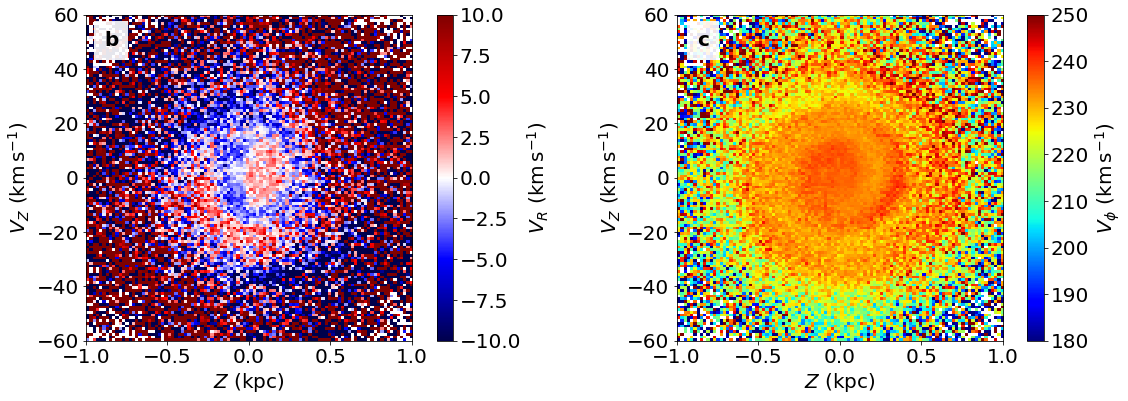}

\caption{{\bf Distribution of stars in the vertical position-velocity plane from \Gaiabf DR2 data.} The panels are for stars in our sample located at $8.24 < R < 8.44 \kpc$. a) Two-dimensional histogram in bins of $\Delta Z=0.01\kpc$ and $\Delta V_Z=0.1\kms$,  with the darkness being proportional to the number of counts.; b) $Z$-$V_Z$ plane coloured as a function of median $\Vr$ in bins of $\Delta Z=0.02\kpc$ and $\Delta V_Z=1\kms$; c) Same as b) but for $\Vp$. }
\label{fig:ZVZ}
\end{figure}

\begin{figure}[H]
\includegraphics[width=0.3\linewidth]{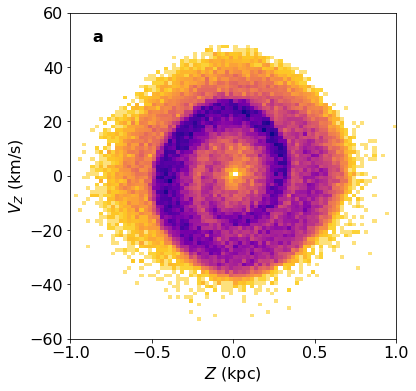}

\includegraphics[width=0.5\linewidth]{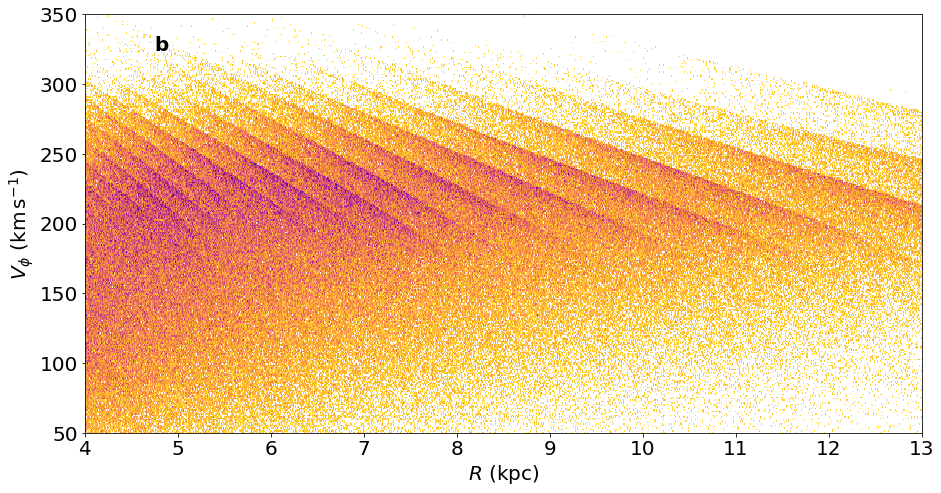}
\includegraphics[width=0.5\linewidth]{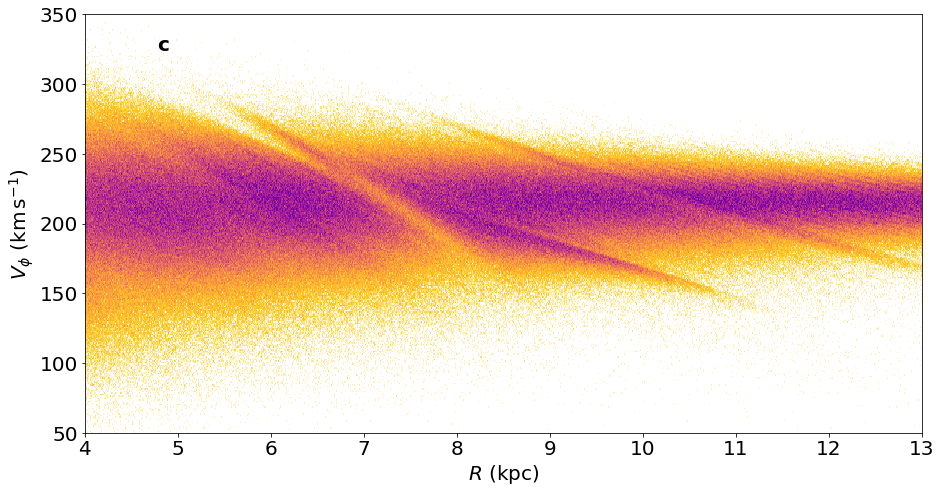}

\caption{{\bf Models of the phase space distribution of the Galaxy disk.} 
a) Modelled spiral shape created in the vertical position-velocity plane as a result of the phase mixing in the evolution of an ensemble of particles for 500 Myr in a Galactic potential, starting from a distribution that is out of equilibrium presumably after a certain perturbation (see Methods);
b) Modelled diagonal ridges created in the distribution of azimuthal velocities as a function of Galactocentric radius  as a result of the phase mixing in the evolution of an ensemble of particles for 1000 Myr in a Galactic potential, starting from a distribution that is out of equilibrium (see Methods); 
c) Same as b) but  for the diagonal ridges created as a result of the effects of the barred potential and its resonant structure (see Methods).}
\label{fig:models}
\end{figure}

\section*{Methods}

\subsection*{Data and samples selection}\label{data}

We used \Gaia sources for which the 6D phase space coordinates can be computed, that is all sources with available 5 parameters astrometric solution (sky positions, parallax and proper motions) and radial velocities. We selected only stars with positive parallaxes $\varpi$ with relative uncertainty smaller than $20\%$, i.e. the ones satisfying $\varpi/\sigma_{\varpi}>5$. This selection is to ensure that $1/\varpi$ is a reasonably good estimator of the distance to the stars\cite{DR2-DPACP-38} but alternatively, we used also Bayesian distances (see below). This sample has 6,376,803 stars and it has been well studied and characterised elsewhere\cite{Katz2018}. The data was obtained directly through the following query in the public \Gaia Archive (\href{https://gea.esac.esa.int/archive/}{https://gea.esac.esa.int/archive/}):

\noindent{\begin{verbatim}
SELECT G.source_id, G.radial_velocity, G.radial_velocity_error, 
G.ra, G.ra_error, G.dec, G.dec_error, G.parallax, G.parallax_error, 
G.pmra, G.pmra_error, G.pmdec, G.pmdec_error, 
G.ra_dec_corr, G.ra_parallax_corr, G.ra_pmra_corr, G.ra_pmdec_corr, 
G.dec_parallax_corr, G.dec_pmra_corr, G.dec_pmdec_corr, 
G.parallax_pmra_corr, G.parallax_pmdec_corr, G.pmra_pmdec_corr
FROM gaiadr2.gaia_source G
WHERE G.radial_velocity IS NOT Null AND G.parallax_over_error>5. 
\end{verbatim}
}

From the 5 parameter astrometric solution and line-of-sight velocities $(\alpha, \delta, \varpi, \mu_{\alpha}^*,\mu_{\delta},V_{los})$ of these stars, we derived distances (as $1/\varpi$), positions and velocities in the cylindrical Galactic reference frame, that is ($R$, $\phi$, $Z$, $\Vr$, $\Vp$, $\VZ$). For convenience, we took $\phi$ positive in the direction of Galactic rotation and with origin at the line Sun-Galactic Centre. For these transformations, we adopted a vertical distance of the Sun above the plane of\cite{Chen2001} $27$ pc, a distance of the Sun to the Galactic centre\cite{Reid2014} $R_{\odot}$ of $8.34$ kpc and a circular velocity at the Sun radius of\cite{Reid2014} $V_\text{c}(R_{\odot})=240$ $\kms$. We assumed a peculiar velocity of the Sun with respect of the Local Standard of Rest of \cite{Schonrich2012} $(U_{\odot},V_{\odot},W_{\odot})=(11.1, 12.24, 7.25)$ $\kms$. Our choice of values gives $(V_\text{c}(R_\odot) + V_{\odot})/R_{\odot}=30.2\kms$\,kpc$^{-1}$, which is compatible with the reflex motion of  Sgr A*\cite{Reid2004}.  
To derive the uncertainties in these coordinates, we propagate the full covariance matrix. The median errors in the $\Vr$, $\Vp$, $\VZ$ velocities are 1.4, 1.5, and $1.0\kms$, respectively, and $80\%$ of stars have errors smaller than 3.3, 3.7, $2.2\kms$ in these velocities.   
The positions in the Cartesian coordinates $X$-$Y$ and $X$-$Z$ of the sample are shown in \edfigref{fig:XYZ}.

For part of our study, we selected from our sample the 935,590 stars located in the solar Galactic cylindrical ring, that is with Galactocentric radius $8.24<R<8.44\kpc$ (dotted lines in \edfigref{fig:XYZ}). For this selection, the median errors in the $\Vr$, $\Vp$, $\VZ$ velocities are 0.5, 0.8, and $0.6\kms$, respectively, and $80\%$ of stars have errors smaller than 1.1, 2.0, $1.3\kms$ in these velocities.

We note that the velocity uncertainties are significantly smaller than the sizes of the substructures detected and that, together with the number of stars in our samples, this is what made possible their detection. Although there are some correlations between the astrometric \Gaia observables\cite{Lindegren2018}, these are not responsible for the correlations and substructure seen in our phase space plots. This is because the stars in our sample are distributed through all sky directions, and the phase space coordinates come from combinations of astrometric measurements and radial velocities, in different contributions depending on the direction on the sky. Besides, the astrometric correlations for our sample are small (smaller than 0.2 in their absolute value for more than 50\% of stars) and this, combined with the small errors, makes their contribution nonsignificant.

Alternatively, we used distances determined through a Bayesian inference method using the existing implementation in TOPCAT\cite{Taylor2005}, taking the mode of the posterior distribution and a prior of an exponentially decreasing density of stars with scale length of 1.35 kpc\cite{Astraatmadja2016}. We found that the differences between this distance determination and the inverse of the parallax are between -2\% and 0.6\% for 90\% of the 6,376,803 stars with $\varpi/\sigma_\varpi>5$, which was expected for small relative errors in parallax. Consequently, the phase space diagrams presented here vary only at pixel level. These panels do not vary 
 even when using the set of 7,183,262 of stars with available radial velocities that include stars with larger parallax errors and stars with negative parallaxes, for which the estimator of the inverse of the parallax would yield unphysical distances. When using another alternative set of Bayesian distances specifically derived for stars from Gaia DR2 with radial velocities using a different prior\cite{McMillan2018}, we found the differences between these distances and the inverse of the parallax to be between -9\% and 5\% for 90\% of the stars, thus slightly larger than before, but again with no noticeable effects on the phase space panels examined here.


\subsection*{Models for the vertical phase mixing}

We first reproduced the spiral shape observed in the $Z$-$\VZ$ plane with the \Gaia DR2 data by using a simple toy model. Often the classic harmonic oscillator is employed to describe the vertical movement of stars in galaxy disks under the epicyclic theory\cite{Bible2008}. However, in this approximation, which is valid only for very small amplitude orbits for which the potential changes little vertically, stars have the same vertical oscillatory frequency $\nu$ and there is no phase mixing, unless orbits at different guiding radius, thus with different frequencies, are considered. Instead, we used an anharmonic oscillator with the potential of \equref{anharmonic}.  
 We took the coefficients $\alpha_0, \alpha_1,\alpha_2$ corresponding to the expansion for small $Z$, derived elsewhere\cite{Candlish2013}, of a Miyamoto-Nagai potential\cite{Miyamoto1975} with values of $a=6.5\kpc$, $b=0.26\kpc$, $M=10^{11}\,\rm{M}_\odot$. These coefficients $\alpha$ depend on Galactocentric radius $R$. 
In this anharmonic potential, the frequencies of oscillation are described by \equref{frequencies}, and thus depend on $R$ and on the amplitude of the oscillation $A$.
 
 Given an initial distribution of stars $Z(t=0)$ and $\VZ(t=0)$, the vertical amplitudes of the orbits can be derived through the conservation of energy and using the fact that at the vertical turn-around point of the orbit ($V_Z=0$),  the (vertical) kinetic energy is null\cite{Candlish2013}. 
Assuming that stars follow a simple harmonic oscillation (but with different frequencies), the movement of the stars with time is described by \equref{harmonic}
\noindent where the initial phase of the stars $\phi_0 \equiv\phi(t=0)$ is obtained from the initial distribution of $Z$ and $V_Z$ and the corresponding amplitudes. 

The phase space evolution described above is shown in the top row of \edfigref{fig:snailmodel}. Initially, the particles followed a Gaussian distribution in $Z(t=0)$ and in $V_Z(t=0)$ with mean and dispersion of $-0.1\kpc$ and $0.04\kpc$, and $-2\kms$ and $1\kms$, respectively. We located all particles at the same Galactocentric radius $R=8.5\kpc$, and thus, they all move under the same functional form of the vertical potential. The initial conditions are shown in \edfigref{fig:snailmodel}a, where we colour-coded the particles according to their period. Following \equref{harmonic}, each star follows a clockwise rotation in the $Z$-$V_Z$ plane. However, they do it at a different angular speed: stars with smaller period located at the closer distances from the mid-plane ($Z=0$) revolve faster than those located at the largest distances from the mid-plane. The whole range of frequencies is what creates, therefore, the spiral shape. \edfigref{fig:snailmodel}b shows the evolution of the system for three initial phases of the time evolution when the spiral shape begins to form. \edfigref{fig:snailmodel}c shows the spiral shape  after $1000\Myr$ of evolution.

In the \Gaia data, we do not see a thin spiral but a thick one, with many of the stars in the volume participating in it. A similar effect was reached with our toy model when we included particles at different radius for which the vertical potential changes and the range of amplitudes/frequencies also changes. In \edfigref{fig:snailmodel} (bottom row) we let the same system evolve as in the top row but starting with initial radius following a skewed normal distribution, which creates a density decreasing with radius as in galaxy disks, with skewness of $10$, location parameter of $8.4\kpc$ and scale parameter of $0.2\kpc$. The spiral structure is now thickened similarly to the data, with higher density of stars at the leading edge of the spiral.


To estimate the time of the phase mixing event from the spiral seen in the \Gaia data (\figref{fig:ZVZ}) using \equref{time}, we 
 needed to locate  two consecutive turns of the spiral and estimate their vertical frequencies from their amplitudes and mean radius. For this, we used \edfigref{fig:aproxlines} which has been colour coded as a function of median guiding radius. This was approximated as $R_g\sim\frac{\Vp R_\odot}{V_c(R_\odot)}$,  under the hypothesis of a flat rotation curve, where we used the values of $R_\odot=8.34\kpc$ and $V_c(R_\odot)=240\kms$ assumed in the coordinate transformation of the data.   In this panel we see that the density gradient across the spiral shape is created by stars with different guiding radius that arrive at the solar neighbourhood due to their different amplitudes of (horizontal) radial oscillation.
 To determine two consecutive turns of the spiral, we focused on stars at the turn-around points ($\VZ=0$) near  
  the leading edges of the spiral. By visual inspection, we determined an approximate range of $Z$ in which the turn around points 
  are located in \edfigref{fig:aproxlines}, concentrating on red colours, for which the spiral is well defined.
  The ranges of the turn around points are marked with vertical lines 
  and listed, together with the middle value, in \edtabref{tab:spiral}. For these turn-around points, the amplitudes are simply $A=Z$ 
  and from the colour bar we note that the average $R_g$ is around $8.2\kpc$. Small changes in this value do not change significantly our final determination of the time of the perturbation.

To estimate the vertical orbital frequencies of these turn-around points, we could not use the toy model presented above since it is valid only for oscillations with small amplitude $A$, in particular smaller than the vertical scale $b$ of the potential ($A<<0.26 \kpc$). Therefore, we took the model of Allen \& Santillan\cite{Allen1991} with updated parameters that fit current estimations such as for the Sun Galactocentric radius and the circular velocity curve \cite{Irrgang2013}. 
We computed the vertical frequency numerically in a grid of different radius and vertical amplitudes by integrating orbits and measuring their vertical periods (\edfigref{fig:freq}a). The vertical frequency can change along the orbits for stars with large eccentricities in the horizontal direction, 
but here for simplicity we put all particles on near circular orbits. We estimated the vertical frequency at each turn-around position directly by interpolating the numbers of \edfigref{fig:freq}a using the estimated values for the amplitude and radius.

Finally, taking each pair of turning points, we obtained an estimation of the time since the perturbation using \equref{time}. As an example, the two turning points (amplitudes) of the left part of the spiral are located at -0.59$\pm9 $ and -0.23$\pm5 \kpc$, respectively. These correspond to vertical frequencies of 0.058$^{+-0.002}_{--0.002}$ and 
0.072$^{+-0.002}_{--0.002}$ rad$\,$Myr$^{-1}$ for $R_g=8.2\kpc$, which gives a time of 461$^{+  183}_{-  105}$ Myr. 
We repeated the same procedure for the second pair of consecutive turning points and also for the edges of the spiral at $Z=0$ (mid-plane points), which have $\VZ=A_{\VZ}\equiv A\nu$, estimating the frequencies by interpolating the values of \edfigref{fig:freq}b. The mid-plane positions are marked as horizontal lines in \edfigref{fig:aproxlines}. All results are summarised in \edtabref{tab:spiral} and \edtabref{tab:spiral2}. The mean of the three time estimations is $510\Myr$ and the minimum and maximum times from the uncertainty ranges are 356 and $856\Myr$. 

We tested the dependence of our time determination on the potential model used by using a different model\cite{McMillan2017}. Compared to our previous model, this one has different shape for the halo, disks and bulge, a different total mass and includes thin and thick disks as well as two gas disks. The frequencies in this  model are on average smaller by 4\% and smaller than 6\% for 90\% of the points in the grid of \figref{fig:freq}a. By repeating the whole process to determine the perturbation time, we obtained $528\Myr$ with minimum and maximum times from the uncertainty ranges of 361 and $899\Myr$, thus, very similar to our previous determination.

We note that our determination is 
subject to several approximations, namely that we used the vertical frequencies of 
 orbits with conditions of circularity on the plane and for certain assumed Galactic potential models, we considered a unique guiding radius, and we took equal initial phases for the turn-around and mid-plane points.

 We finally run a simulation (\figref{fig:models}a) by integrating 100,000 test particle orbits in the updated Allen \& Santillan model with initial vertical positions and velocities following Gaussian distributions centred at $Z=-0.4\kpc$ and  $\VZ=-5\kms$ and dispersions of $0.15\kpc$ and $2.\kms$, respectively. Horizontally in the disk plane, they were distributed following a skewed normal in radius $R$ with scale parameter of 0.8 kpc, location parameter of 8, skewness of 10, and all particles at an azimuthal angle $\phi=0$. 
The horizontal velocities were set to 0 for the radial component and to the circular velocity at the particle's radius  and $Z=0$ for the azimuthal one. These are initial conditions of circularity on the Galactic plane but not necessarily for orbits with large excursions in $Z$. The particles orbits were integrated forwards in time for $500\Myr$ as estimated from the data.
We note that this is not meant to be a fit to the data since we have not explored all possible initial configurations that could lead to a similar spiral shape. We see, though, that an initial distribution  asymmetric in $Z$ with most particles located at positive or negative $Z$ is required to obtain a single spiral instead of a symmetrical double one. 

\subsection*{Model for the horizontal phase mixing}

 We used a simple toy model to reproduce the lines observed in the $\Vp-R$ plane (\figref{fig:models}b).
This model is built by integrating orbits in the Galactic potential of Allen \& Santillan\cite{Allen1991} with updated parameters that fit current estimations such as for the Sun Galactocentric radius or the circular velocity curve \cite{Irrgang2013}. We used as initial conditions a set of test particles distributed in Galactocentric radius according to a skewed normal distribution with skewness of $10$, location parameter of $4\kpc$ and scale parameter of $6\kpc$. The azimuthal angle was fixed at 0. For simplicity, all particles were put at the mid-plane with null vertical velocities. The radial and azimuthal velocities were initialised, respectively, following Gaussian distributions centred at 0 with dispersion of $40\kms$ and centred at the circular velocity at the particular radius with a dispersion of $30\kms$. The particle orbits were computed for $1000\Myr$. 


\subsection*{Model for the horizontal resonances}

The model of  \figref{fig:models}c is from a test particle simulation of orbits integrated in a Galactic potential model including a bar\cite{RomeroGomez2015}. The axisymmetric part of the potential was the  
Allen \& Santillan model\cite{Allen1991}. The Galactic bar potential was built using Ferrers ellipsoids\cite{Ferrers1877} oriented 
with its semi-major axes at $20 \deg$ from the Sun-Galactic centre line, and its pattern speed was set to 50 $\kmskpc$, which corresponds to a period of about $120\Myr$. 
The simulation consisted of 68 million test particles with an initial radial velocity dispersion of $30\kms$ 
at the Solar radius.
Their orbits were first integrated in the axisymmetric potential model for $10\Gyr$ until they were approximately fully phase mixed. Next, the bar potential was grown in $T_{\rm{grow}}\equiv4$ bar rotations. 
More details on the bar potential, initial conditions and integration procedure are specified elsewhere\cite{RomeroGomez2015}. Here we used the final conditions after $T_{\rm grow}$ ($\sim500\Myr$) and 8 additional bar rotations ($\sim1000\Myr$). From all the particles in the simulation, we used only the ones located in a range of $10\deg$ in azimuthal angle centred on the Sun, similar to the our data sample. This selection contains 2,009,791 particles.

\renewcommand{\refname}{Additional references}

\subsection*{Data availability statement}

The datasets used and analysed for the current study are derived from the data available in the public \Gaia Archive, [\href{https://gea.esac.esa.int/archive/}{https://gea.esac.esa.int/archive/}]. \b{The Bayesian distances for the Gaia sources with radial velocity\cite{McMillan2018} are available at \url{http://www.astro.lu.se/~paul/GaiaDR2_RV_star_distance.csv.gz}.} The \b{rest of} datasets and toy models generated and/or analysed are available from the corresponding author on reasonable request.
 
\subsection*{Code availability statement}
\b{We have made use of standard data analysis tools in Python environment. The codes to generate the toy models,  simulations and compute orbital frequencies are available from the corresponding author on reasonable request.
The code used to compute the orbits for the McMillan 2017 potential\cite{McMillan2017} is available at \href{https://github.com/PaulMcMillan-Astro/GalPot}{https://github.com/PaulMcMillan-Astro/GalPot}. The code to compute Bayesian distances from parallaxes is available in the TOPCAT platform\cite{Taylor2005}.}




\setcounter{figure}{0}
\section*{Extended Data}

\renewcommand*{\tablename}{Extended Data Table}
\renewcommand*{\figurename}{Extended Data Figure}

\subsection*{Extended Data Tables}

 \begin{table}[H]
\caption{\label{tab:spiral}{\bf Time estimations from the turn-around points of the spiral. }}
\centering
 \label{tab:spiral}
\includegraphics[width=0.3\linewidth]{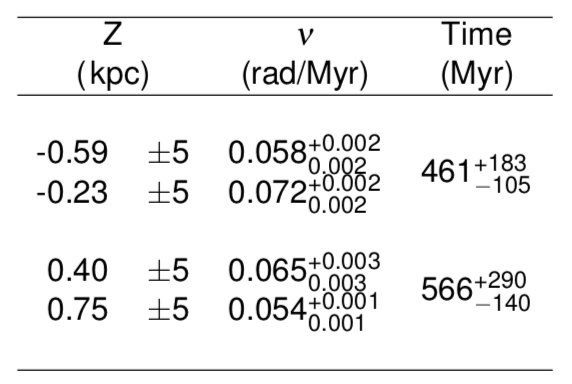}
\begin{flushleft}
The first column indicates the vertical position of the turn-around points, which are equal to the amplitude of the orbits except for the sign, and the estimated uncertainty ranges.The following columns are the frequencies corresponding to these amplitudes, and the starting times of the phase mixing process corresponding to each pair of consecutive spiral turns.
\end{flushleft}
\end{table}

 \begin{table}[H]
\caption{\label{tab:spiral2}{\bf Time estimations from the mid-plane points of the spiral.} }
\centering
 \label{tab:spiral2}
\includegraphics[width=0.3\linewidth]{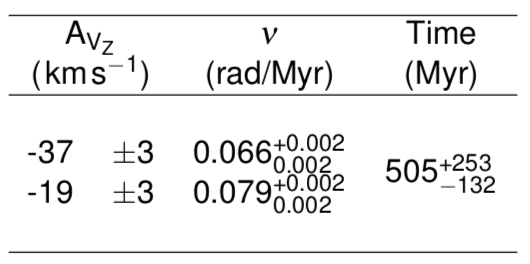}
\begin{flushleft}
The first column indicates the vertical velocity at the mid-plain passages, which are equal to the velocity amplitudes of the orbits except for the sign, and the estimated uncertainty ranges.The following columns are the frequencies corresponding to these amplitudes, and the starting time of the phase mixing process corresponding to the pair of consecutive spiral turns.
\end{flushleft}
\end{table}

\subsection*{Extended Data Figures}

\begin{figure}[H]
\includegraphics[width=1.\linewidth]{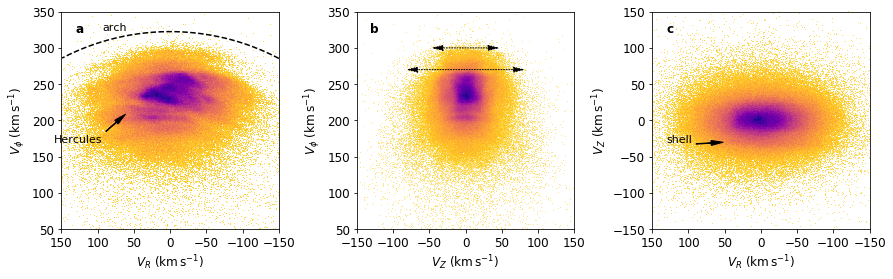}
\caption{{\bf Distribution of velocities in the solar Galactocentric radius from \Gaiabf DR2 data.} Two-dimensional histograms of \b{combinations of} radial, azimuthal and vertical Galactic cylindrical velocities for the stars in our sample located at $8.24 < R < 8.44 \kpc$, in bins of $1\kms$. $\Vr$ and  $\Vp$ are positive towards the Galactic anti-centre and the Galactic rotation direction, respectively. The darkness is proportional to the number of counts. 
\b{a)} Although the bimodality seen in \b{this} panel, separating the Hercules stream from the rest of the distribution was known\cite{Eggen1958,Blaauw1970}, \b{as well as some other elongated structures in this velocity projection
\cite{Dehnen1998,Skuljan1999,Antoja2008}} the numerous \b{and thin} arches in this panel are a new phenomenon revealed by {\it Gaia} \b{elsewhere\cite{Katz2018}}. The semi-circular dotted line in \b{this} panel marks an arbitrary  line of constant kinetic energy in the plane $E_k=\frac{1}{2}(\Vr^2+\Vp^2)$, \b{as predicted for the substructure generated in horizontal phase mixing\cite{Minchev2009,Gomez2012a}.}
\b{b)} This panel depicts a rather boxy appearance, where the extent of the  arches in $V_Z$, varies with their $\Vp$ (arrows), \b{likely created by the correlation between the spiral shape and the $\Vp$ velocities of \figref{fig:ZVZ}c}.
\b{c) While some velocity asymmetries where noticed before in the $\Vp-\VZ$ projection\cite{Dehnen1998} and attributed to the Galaxy warp,} 
  the \b{sharp} shell-like features involving the $V_Z$ velocities, \b{especially at $\VZ\sim-30\kms$ and $\VZ\sim25\kms$, are unveiled here for the first time. These shells are different projections of the ``snail shell'' pattern of \figref{fig:ZVZ}a}. }
\label{fig:VrVphiVzSN}
\end{figure}

\begin{figure}[H]
\centering
\includegraphics[width=0.4\linewidth]{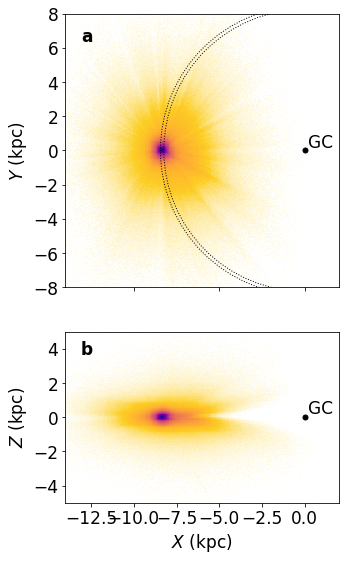}
\caption{{\bf Distribution of stars in our sample in the disk of the Galaxy.}  Two dimensional histograms with bins of $0.05\kpc$ in the $X$-$Y$ and $X$-$Z$ projections. The dotted lines mark the selection of stars in the solar Galactic ring of between radius of $[8.24,8.44]\kpc$. The Sun is located at \b{$(X,Y,Z)=(-8.34,0.,0.027)\kpc$} and the Galactic Centre is marked with a black dot.}
\label{fig:XYZ}
\end{figure}



\begin{figure}[H]
\centering
\includegraphics[width=00.9\linewidth]{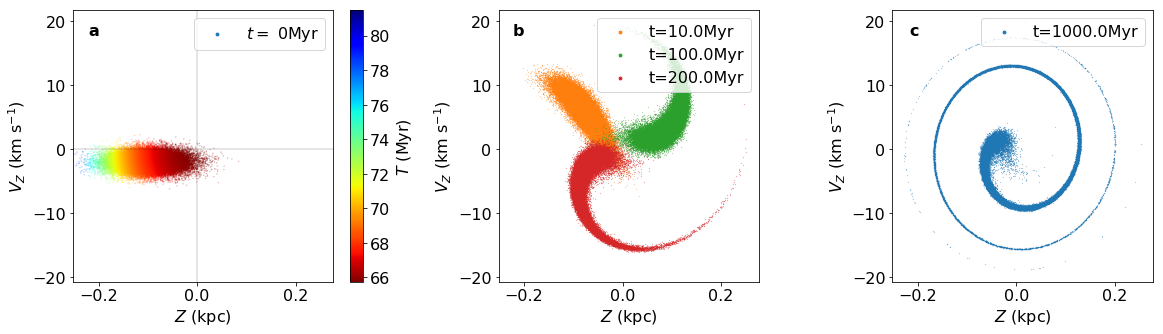}

\includegraphics[width=00.9\linewidth]{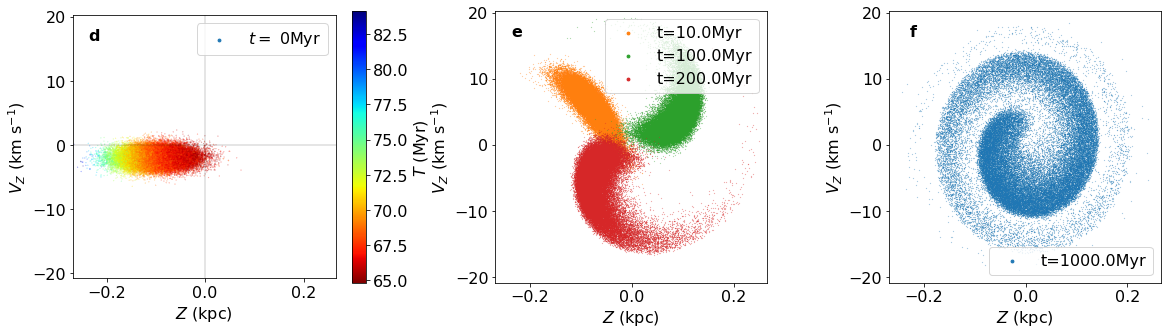}
\caption{{\bf Spiral shape created in the phase space evolution under an anharmonic potential.}  a-b-c) Phase space evolution \b{at different times (t=0,10,100,200,1000 Myr)} for an ensemble of particles at a fixed Galactocentric radius of $R=8.5\kpc$ with an initial Gaussian distributions in $Z(t=0)$ with mean of $-0.1\kpc$ and dispersion of $0.04\kpc$ and in $V_Z(t=0)$ with mean of $-2\kms$ and dispersion of $1\kms$. d-e-f) Same as a-b-c) but for a skewed normal distribution of initial radius with skewness of $10$, location parameter of $8.4\kpc$ and scale parameter of $0.2\kpc$. In both rows, the evolution is the one of an anharmonic oscillator derived from the expansion of a Miyamoto-Nagai disk for small $Z$. \b{In panels a) and d) the stars are colour coded by vertical period.} }
\label{fig:snailmodel}
\end{figure}

\begin{figure}[H]
\centering
\includegraphics[width=0.49\linewidth]{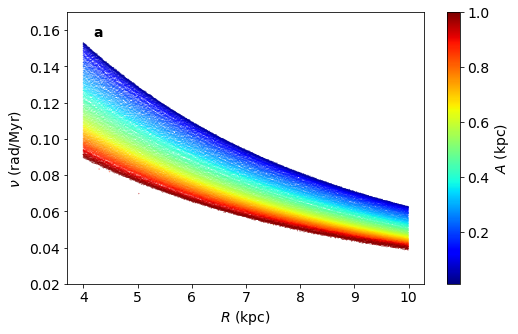}
\includegraphics[width=0.49\linewidth]{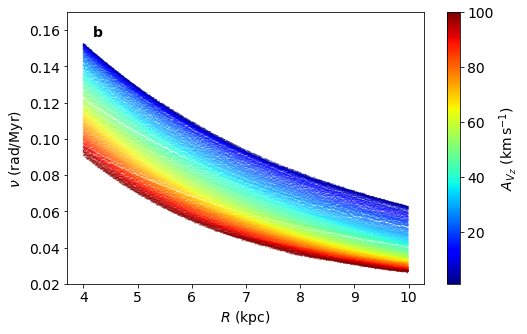}
\caption{{\bf  Vertical frequency for orbits in a Galaxy model potential.} Frequencies as a function of Galactocentric radius $R$ \b{computed in the updated Allen \& Santillan model \cite{Irrgang2013}} a) colour coded by vertical amplitude of the orbits and b) colour coded by vertical velocity amplitude of the orbits. }
\label{fig:freq}
\end{figure}

\begin{figure}[H]
\centering
\includegraphics[width=0.49\linewidth]{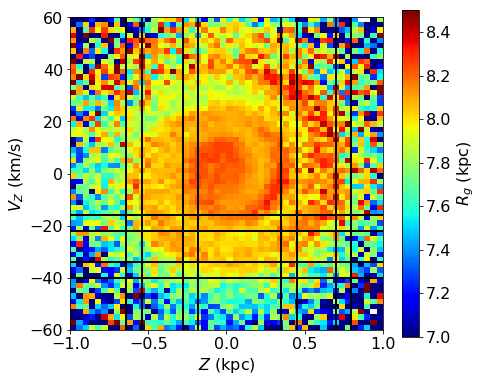}
\caption{{\bf Position of the shells in the distribution of stars in the vertical position-velocity plane.} 
$Z$-$V_Z$ plane coloured as a function of median guiding radius $R_g$ for stars at Galactocentric radius of [8.24,8.44] $\kpc$ in bins of $\Delta Z=0.02(\kpc)$ and $\Delta V_Z=1\kms$ with horizontal and vertical lines showing the approximate locations of the observed shells \b{(turn-around and mid-plane points)}.  }
\label{fig:aproxlines}
\end{figure}


\end{document}